\documentclass[a4paper,11pt]{article}
\usepackage{pos}
\usepackage{multirow}
\usepackage{subcaption}
\usepackage{enumitem}

\title{Modeling intrinsic time-lags in flaring blazars in the context of Lorentz Invariance Violation searches}
 \ShortTitle{Modeling intrinsic time-lags in flaring blazars for LIV searches}

\author*[1,2]{Christelle Levy}
\author[2]{Hélène Sol}
\author[1]{Julien Bolmont}

\affiliation[1]{Sorbonne Université, CNRS, IN2P3, Laboratoire de Physique Nucléaire et de Hautes Energies (LPNHE), 4 place Jussieu, F-75252 Paris, France}

\affiliation[2]{Laboratoire Univers et Théorie (LUTh), Observatoire de Paris-Meudon, 5 Place Jules Janssen, 92190 Meudon, France}

\emailAdd{clevy@lpnhe.in2p3.fr}
\emailAdd{helene.sol@obspm.fr}

\abstract{Some Quantum Gravity (QG) theories, aiming at unifying general relativity and quantum mechanics, predict an energy-dependent modified dispersion relation for photons in vacuum leading to a Violation of Lorentz Invariance (LIV). One way to test these theories is to monitor TeV photons time-of-flight emitted by distant, highly energetic and highly variable astrophysical sources such as flaring active galactic nuclei. Only one time-lag detection was reported so far. We have recently shown however that significant intrinsic time-lags should arise from \textit{in situ} blazar emission processes at TeV energies and should consequently interfere with LIV searches.

In this contribution we will review how intrinsic time delays and LIV-induced propagation effects can simultaneously impact blazars' observed spectral energy distributions and lightcurves. Using a time-dependent approach, we provide predictions on both contributions for various cases in the frame of a standard one zone synchrotron-self-Compton (SSC) model. We will also introduce hints and methods on how to disentangle intrinsic time delays from extrinsic ones in order to highlight LIV effects.}

\FullConference{37$^{\rm{th}}$ International Cosmic Ray Conference (ICRC 2021)\\
		July 12th -- 23rd, 2021\\
		Online -- Berlin, Germany}

\begin{document}
\maketitle

\section{Introduction}
Although some quantum gravity (QG) models have been considered as convincing approaches to the unification of general relativity (GR) and quantum mechanics (QM), it has also been notoriously difficult to extract observable predictions that can be used to test them. Some models (see \citep{Mavromatos2010,Gamibini1999}) predict departures from Lorentz invariance through violations (noted LIV for Lorentz Invariance Violation) expected to occur at energies approaching the Planck scale ($E_P = \sqrt{\hbar c^5 / G}\, \simeq 10^{19}$~GeV) where GR and QM should compete, while retaining the symmetry at lower energies. Amongst other effects, LIV appears as a modified dispersion relation (MDR) of photons in vacuum inducing an energy-dependent velocity, the well-established velocity $c$ becoming a low energy limit \citep{Amelino1998}. 

A strategy currently developed to search for LIV signatures is to look for energy dependent time delays in the gamma-ray signal coming from remote and variable cosmic sources such as blazars at TeV energies \citep{Amelino1998, Liberati2006}. However, \textit{in situ} time delays can also be generated from the radiative process at work in the sources themselves. In case of positive detection, it will be necessary to be able to distinguish any LIV propagation effect from sources' intrinsic effects. One way to address this issue is the study of these intrinsic effects through the modeling of sources emission mechanisms we will present here.

We first introduce a time dependent model of blazar relying on a standard one-zone synchrotron self Compton scenario, and review the study of intrinsic time delays and their properties. Then, we present a multi-frequency study together with methods relying on the euclidian distance and hardness-intensity diagrams that can help discriminate intrinsic from LIV-induced effects.

\section{Time dependent modeling of blazar flares}

To study intrinsic delays, the ones that are inevitable and necessarily arise from emission mechanisms, we build a "minimalist" model based on the most fundamental processes needed to generate a flare: acceleration and radiative cooling. We make use of a standard SSC model widely applied to explain BL Lac flares \citep{Romero} and describe the emitting zone with a single spherical homogeneous bulk of dense leptonic plasma, the surrounding jet medium being neglected. 

The evolution of the lepton number density $N(t,\gamma)$ is used to generate the blazar's emission from infrared to gamma-rays and track its temporal evolution, described by the following equation:
\begin{equation}
\label{eq:ED}
\frac{\partial N(t,\gamma)}{\partial t} = \frac{\partial}{\partial \gamma} \left \{ \left[ C_{rad}(t)\gamma^2 - C_{acc}(t)\gamma \right] N(t,\gamma) \right \}
\end{equation}
where $N(t,\gamma)$ is given in $\mathrm{cm}^{-3}$. $C_{acc}(t) \propto A_0 \left( \frac{t_0}{t} \right)^{m_a}$ and $C_{rad}(t) \propto B_0 \left( \frac{t_0}{t} \right)^{m_b}$ account for acceleration and radiative processes respectively, with parametrisations introduced for the acceleration and magnetic field. $A_0$ and $B_0$ are the initial amplitudes given in $\mathrm{s}^{-1}$ and mG, and $m_a$ and $m_b$ are the evolution indices. $t_0 = \frac{R_0}{c/\sqrt{3}}$ is the characteristic evolution time expressed in seconds, taken as the time needed for a sound wave to propagate through a bulk of radius $R_0$.

Following \cite{Katarzynski2001}
, we introduce the term $\eta = U_{B}(t)/U_{IC}(t)$, ratio of synchrotron to inverse Compton energy densities $U_{B}(t)$ and $U_{IC}(t)$ and approximate the radiative loss term such that:
\begin{equation}
\label{eq:Crad}
C_{rad}(t,\gamma) = \frac{4\sigma_T}{3m_ec} U_B(t) \left( 1 + \frac{1}{\eta} \right).
\end{equation}
An analytical solution to equation~\ref{eq:ED} is then fully determined when an initial condition is provided. We impose a power law with a minimum Lorentz factor $\gamma_{min}$ for leptons and a sharp cut-off at high Lorentz factor that reads
\begin{equation}
\label{eq:ED_0}
N(t_0,\gamma) = N_0 \gamma^{\alpha} \left[ 1 - \left( \frac{\gamma}{\gamma_{cut}(t_0)} \right)^{\alpha+2} \right],
\end{equation}
where $N_0$ is the initial lepton number density, $\gamma_{cut}(t_0)$ is the imposed Lorentz boost cut-off and $\alpha$ is the power law index. Acceleration can then boost leptons up to an upper bound $\gamma_{sup}$.

\begin{figure}
\vspace{-0.25in}
\begin{minipage}[t]{0.32\textwidth}
\centering
\includegraphics[width=1\linewidth]{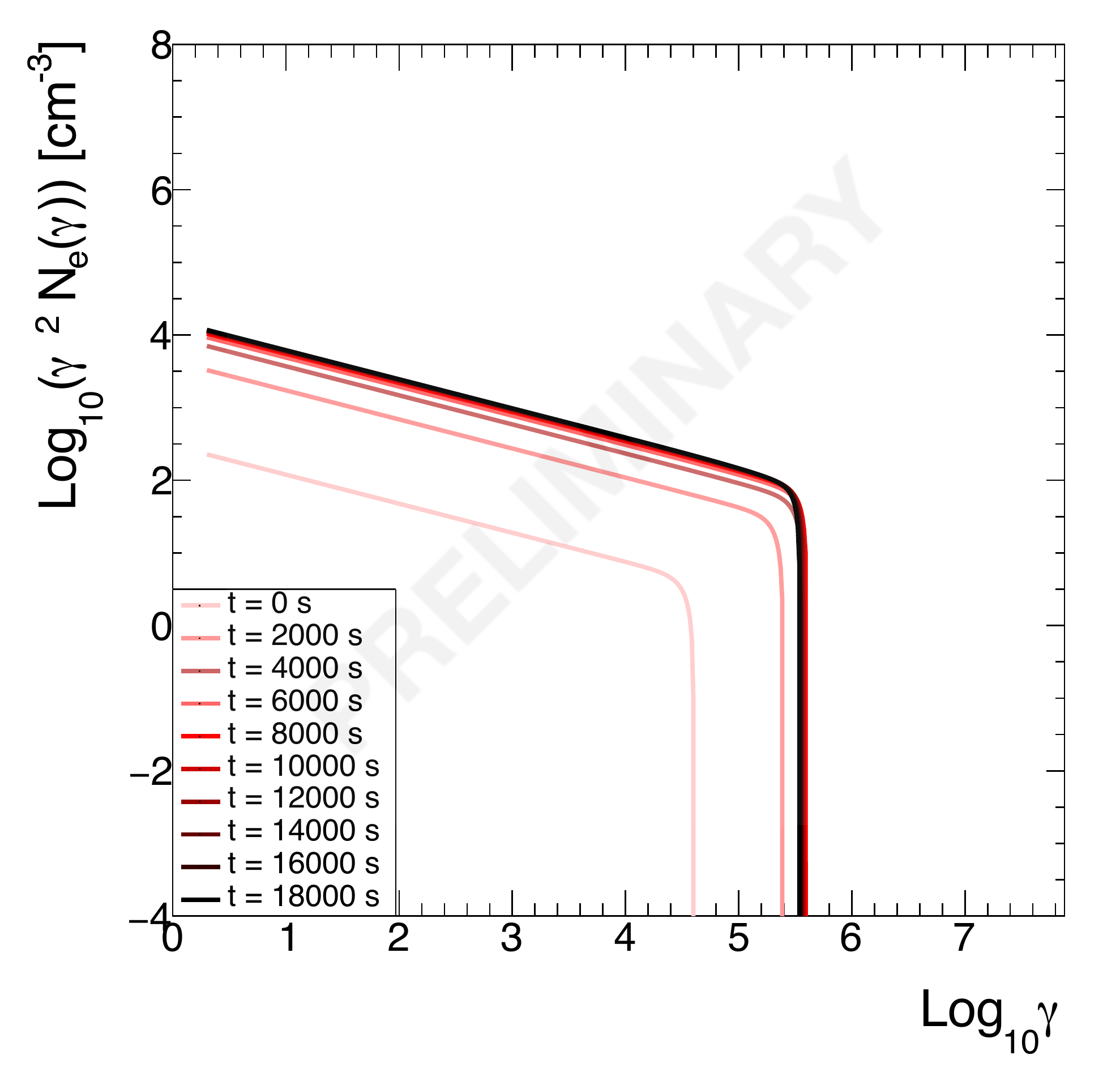}
\vspace{-0.25in}
\subcaption{Temporal evolution of the lepton spectrum.}
\label{fig:ES}
\end{minipage}\hfill
\begin{minipage}[t]{0.32\textwidth}
\centering
\includegraphics[width=1\linewidth]{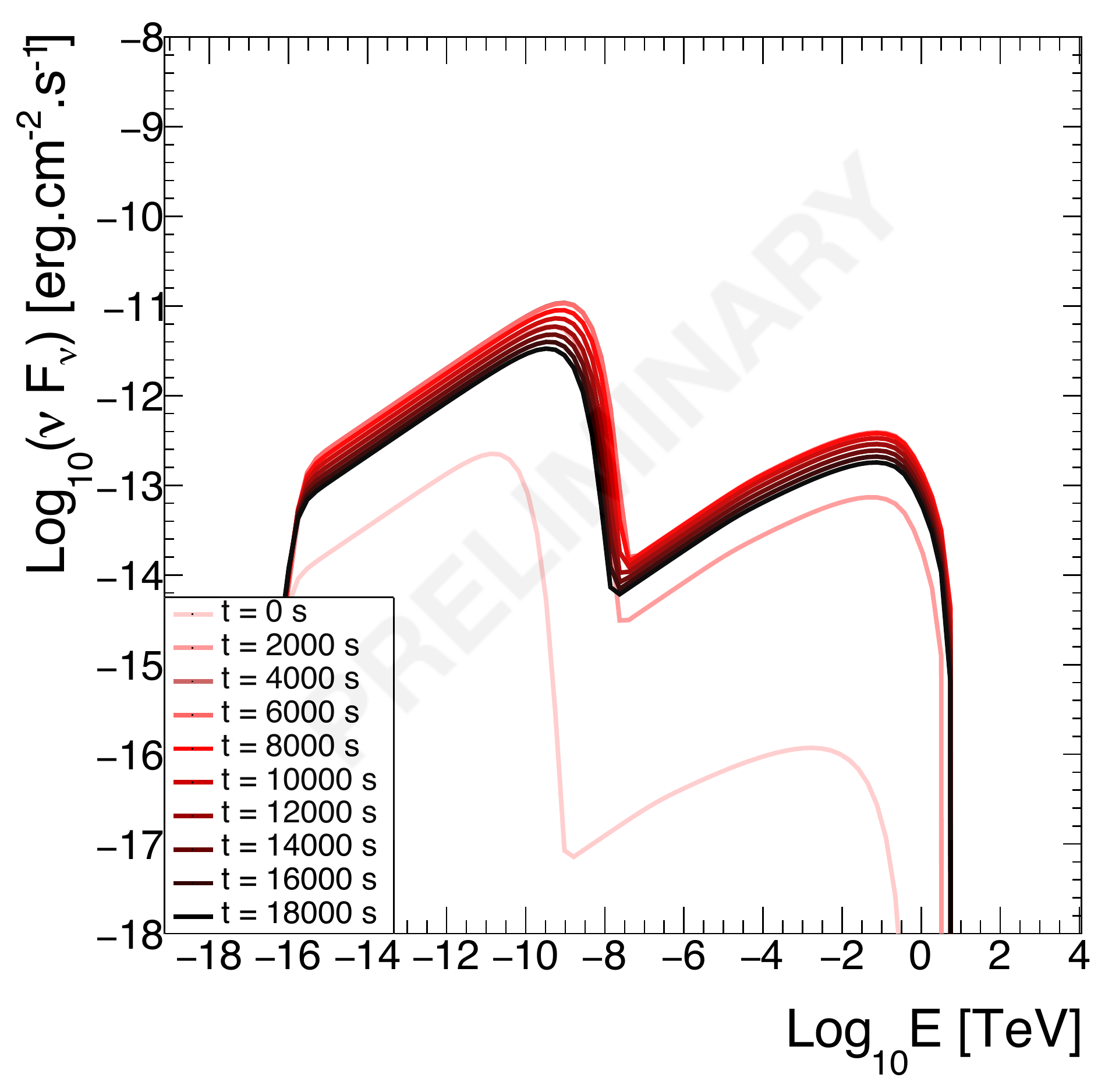}
\vspace{-0.25in}
\subcaption{Temporal evolution of the spectral energy distrubtion (SED).}
\label{fig:SED}
\end{minipage}\hfill
\begin{minipage}[t]{0.32\textwidth}
\centering
\includegraphics[width=1\linewidth]{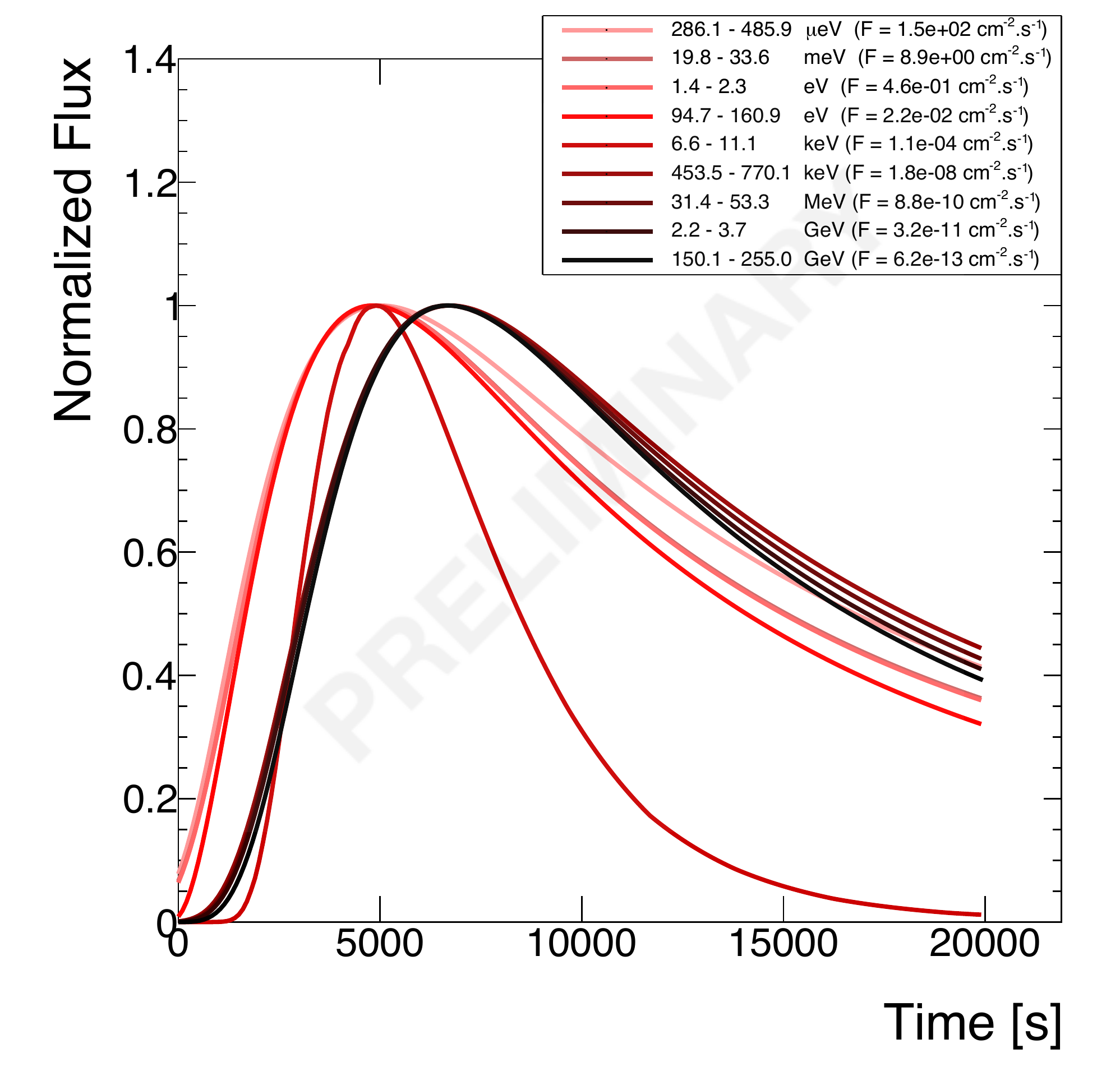}
\vspace{-0.25in}
\subcaption{Light curves in different spectral bands.}
\label{fig:LC}
\end{minipage}
\caption{Example of temporal evolution and energy distributions of blazar observables.} 
\vspace{-0.15in}
\label{fig:observables}
\end{figure}

From the analytical solution, we build and deduce lepton spectra (Figure~\ref{fig:ES}) and their evolution over time, and obtain spectral energy distributions (SED) (Figure~\ref{fig:SED}) and light curves (Figure~\ref{fig:LC}). Two characteristic bumps of SEDs, governed by synchrotron emission at low energy and inverse Compton (IC) emission at high energy, define two energy domains - we define them as synchrotron and IC domains - which will be extensively used later on. The three observables have been generated here for a reference set of parameters based on the properties of the archetypal BL Lac Mrk 501 with a corrective term to account for gamma-ray absorption due to the extragalactic background light (EBL). The same parameters are used to generate all the following results (shown with black distributions), unless specified otherwise.


\section{Intrinsic time delays}
\subsection{Extraction}
Each light curve at energy $E$ can be characterised by a typical arrival time of photons $t_E$, chosen as the time at which the light curve is peaking. To estimate temporal delays between arrival times at energy $E=h\nu_{obs}$, we define a reference arrival time $t_{ref}$ at a specific energy $E_{ref}$. The time delays are then
\begin{equation}
\Delta t_{E} = t_{E} - t_{ref}.
\end{equation}

\subsection{Properties}
First analysed by Perennes et al. \citep{Perennes}, intrinsic delays navigate between two regimes in the SSC scenario, either increasing or decreasing with energy. These two trends arise from an imbalance between the acceleration and the cooling processes, where one process dominates the other. We identify two main regimes,
\begin{itemize}
\item \textbf{Increasing regime:} the acceleration is relatively slow and leptons need time to reach the highest energies. The low energy flares start to be emitted while high energy leptons do not exist yet. Combined with the magnetic field decay, low energy light curves decay before high energy ones leading to increasing time delays.
\item \textbf{Decreasing regime:} the acceleration is relatively fast compared to cooling processes such that leptons are quickly accelerated to their highest energy. Low and high energy photons can be emitted together. Since synchrotron losses are proportional to the square of the lepton energy $P_{ssc} \propto E_l^2$, high energy light curves decay faster than low energy ones leading to decreasing delays.
\end{itemize}
%
%

A third trend can arise between the two others when delays do not vary with eneregy and defines the so-called "flat regime". The reference parameters 
have been fine tuned to reproduce this special case. Indeed, no significant time delay has been detected yet, with the notable exception of one case \citep{Mkn501} such that the flat regime better reflects actual observations. Varying any parameter would immediatly send delays into one of the other regimes, the flat regime being very unstable.

\begin{figure}
\vspace{-0.25in}
\begin{minipage}[t]{0.32\textwidth}
\centering
\includegraphics[width=1\linewidth]{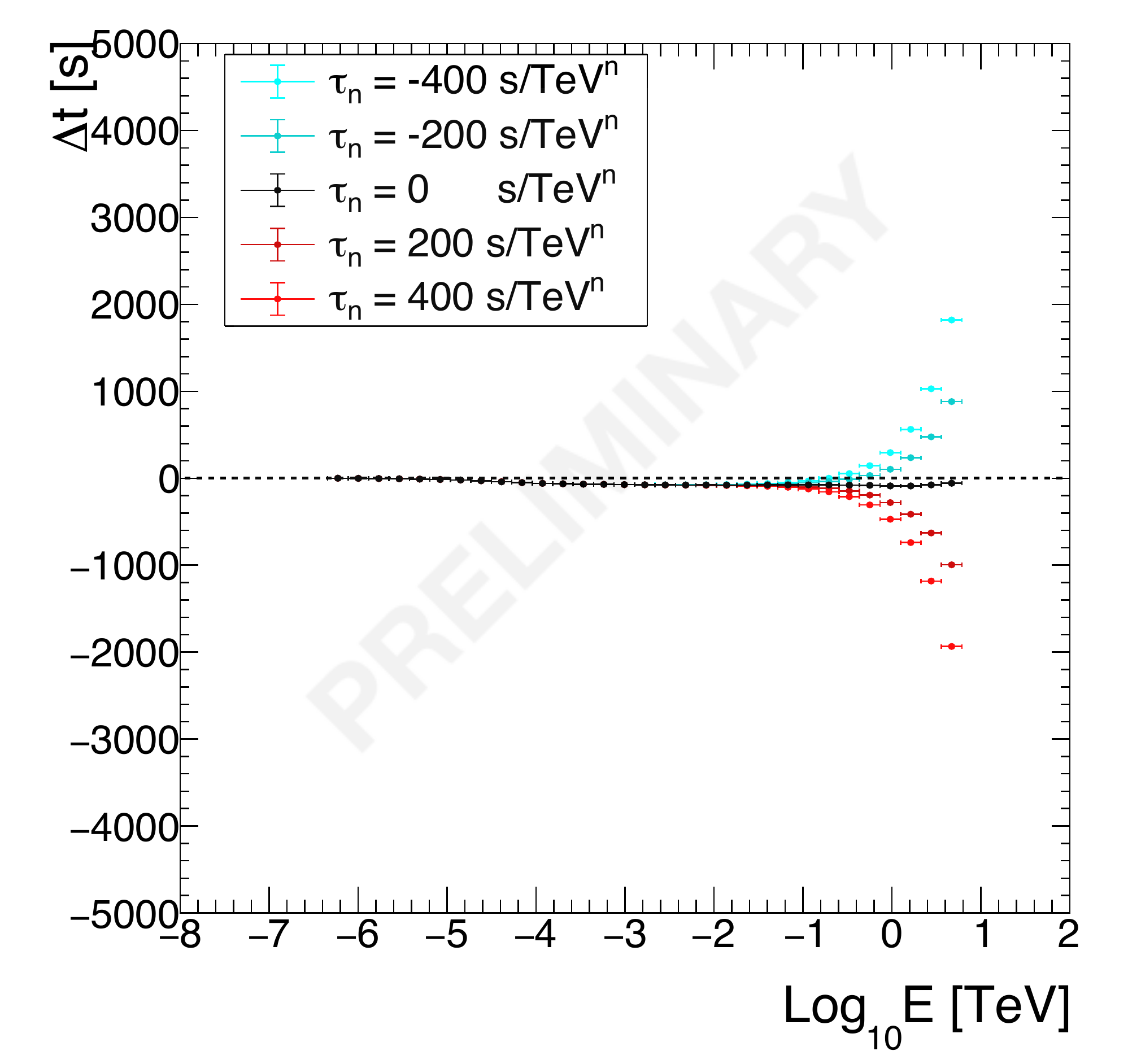}
\vspace{-0.35in}
\subcaption{Flat regime.}
\label{fig:LIVevol_flat1}
\end{minipage}\hfill
\begin{minipage}[t]{0.32\textwidth}
\centering
\includegraphics[width=1\linewidth]{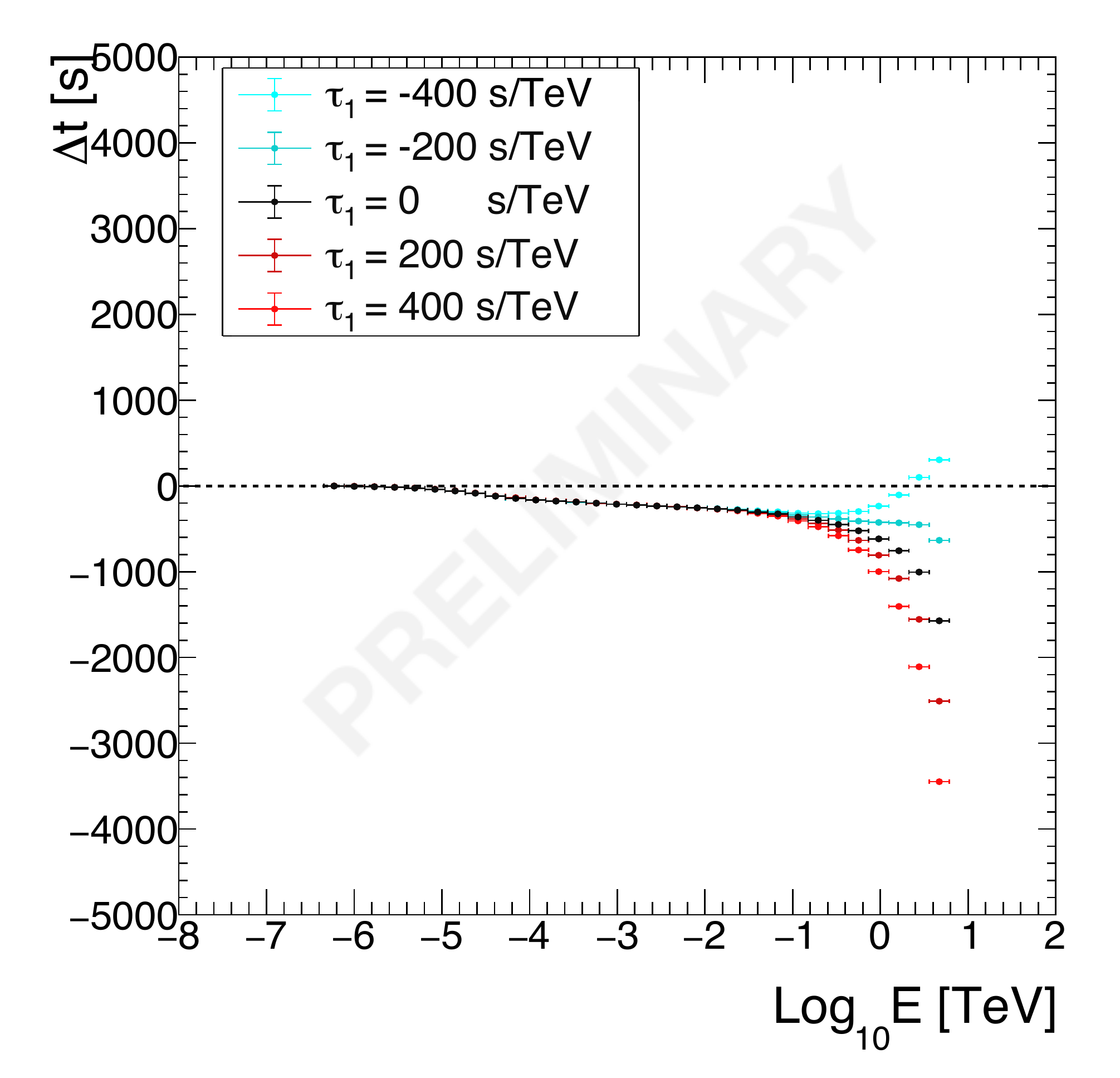}
\vspace{-0.35in}
\subcaption{Decreasing regime.}
\label{fig:LIVevol_decr1}
\end{minipage}\hfill
\begin{minipage}[t]{0.32\textwidth}
\centering
\includegraphics[width=1\linewidth]{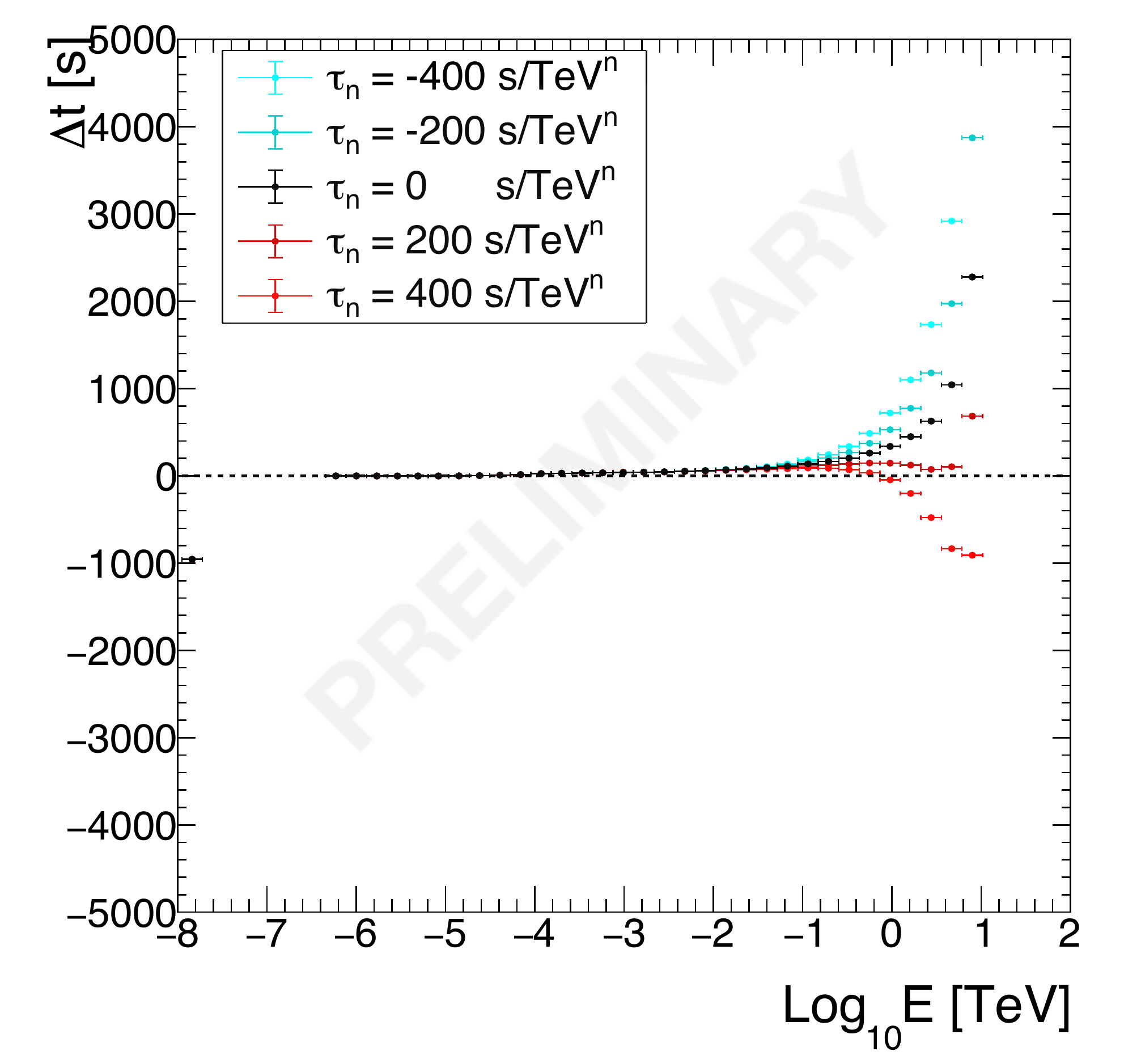}
\vspace{-0.35in}
\subcaption{Increasing regime.}
\label{fig:LIVevol_incr1}
\end{minipage}
\vspace{-0.1in}
\caption{Intrinsic delays with various LIV contribution with $E_{ref}=1$ MeV.} 
\label{fig:LIVevol}
\end{figure}

\subsection{LIV injection}
In order to find a way to discriminate between intrinsic and LIV induced time delays, we introduce LIV effects in the flare description. In the standard approach, they are only dependent on the distance of the source and the simplest way to account for them is to shift light curves by a quantity dependent on their energy:
\begin{equation}
\label{eq:int+LIV}
F_{E_{LC}}(t) \longrightarrow F_{E_{LC}}(t + \tau_n E_{LC}^n).
\end{equation}
where $E_{LC}$ is the mean value of the energy band on which a light curve is defined, $n$ is the LIV correction order, and $\tau_n$ is the LIV term kept as a free parameter expressed in $\mathrm{s/TeV^n}$ which can take positive or negative values leading to subluminal and superluminal velocities respectively. For the sake of illustration, we only consider here linear corrections($n=1$).

As LIV contribution depends on energy, $\tau_n > 0$ tends to send IC delays in the increasing regime and $\tau_n < 0$ in the negative one. Therefore, LIV either amplifies time delays when both intrinsic and LIV effects impose the same regime, or reduce time delays when the two effects impose opposite regimes. For $\tau_n$ big enough, LIV effects can even change the regime in the IC domain as shown in Figure~\ref{fig:LIVevol}. 

\section{Multi-frequency study}
In SSC scenarii the IC emission depends on the synchrotron one and SED bumps evolve together in a rather similar fashion. This similarity between energy bands also appears in light curves and intrinsic time delays. Since LIV effects only contribute at high energies (i.e. in the IC domain), finding a relationship between sets of intrinsic delays arising in the synchrotron and IC domains may allow us to discriminate between intrinsic and LIV induced time delays.

\begin{figure}
\vspace{-0.25in}
\begin{minipage}[t]{0.32\textwidth}
\centering
\includegraphics[width=1\linewidth]{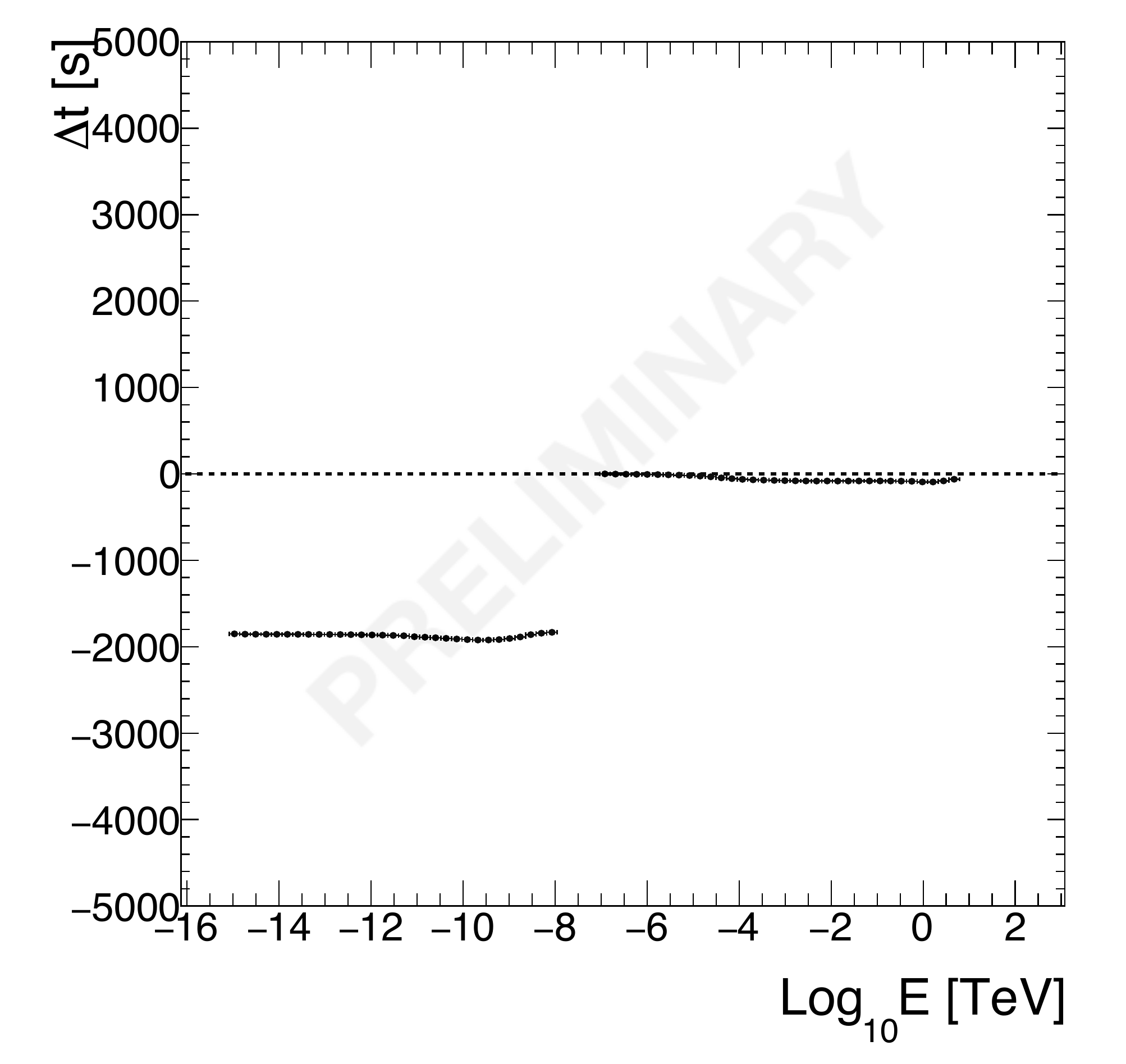}
\vspace{-0.35in}
\subcaption{$B_0=8.7$ mG (flat).}
\label{fig:flat}
\end{minipage}\hfill
\begin{minipage}[t]{0.32\textwidth}
\centering
\includegraphics[width=1\linewidth]{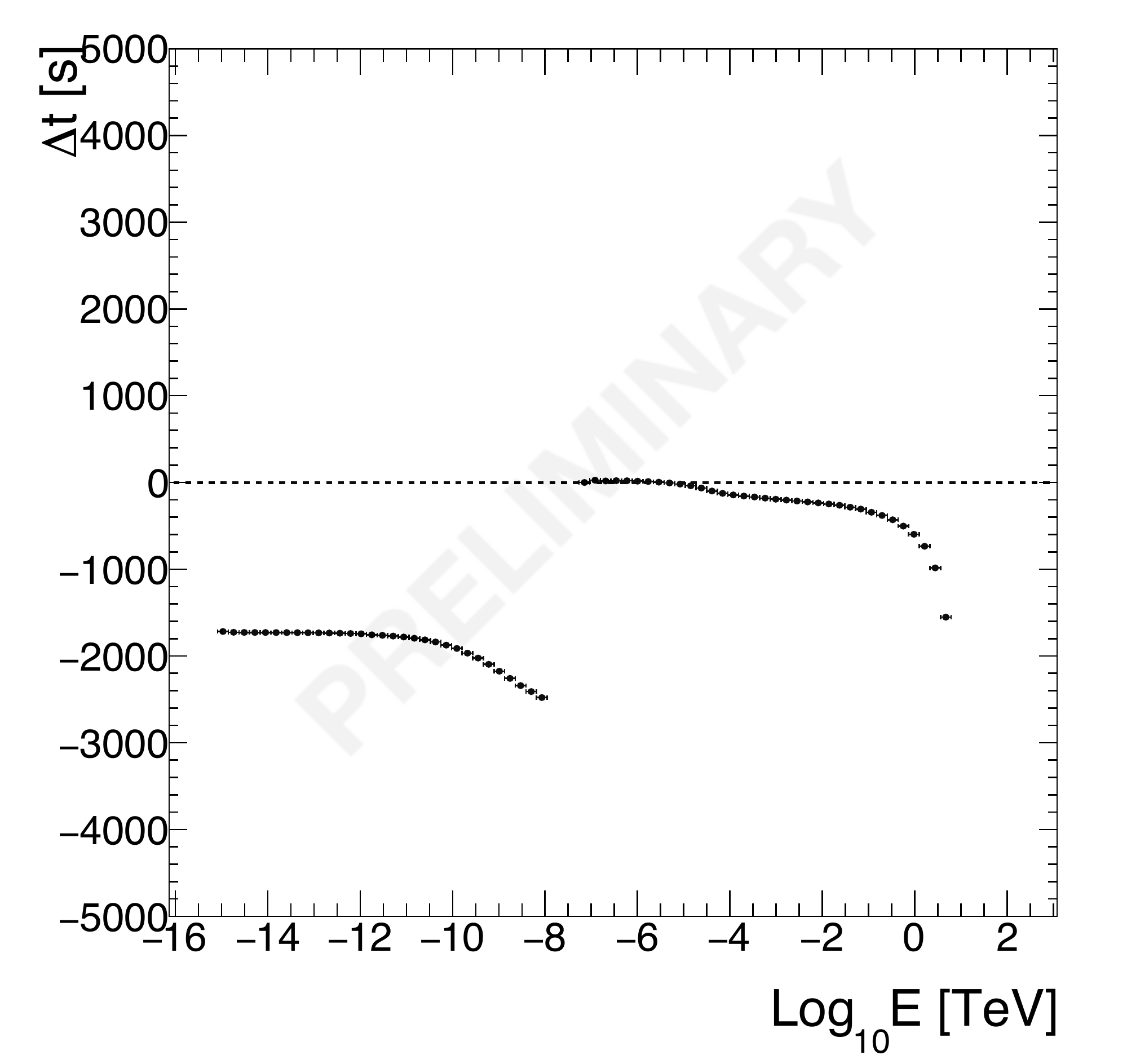}
\vspace{-0.35in}
\subcaption{$B_0=13$ mG (decreasing).}
\label{fig:decr}
\end{minipage}\hfill
\begin{minipage}[t]{0.32\textwidth}
\centering
\includegraphics[width=1\linewidth]{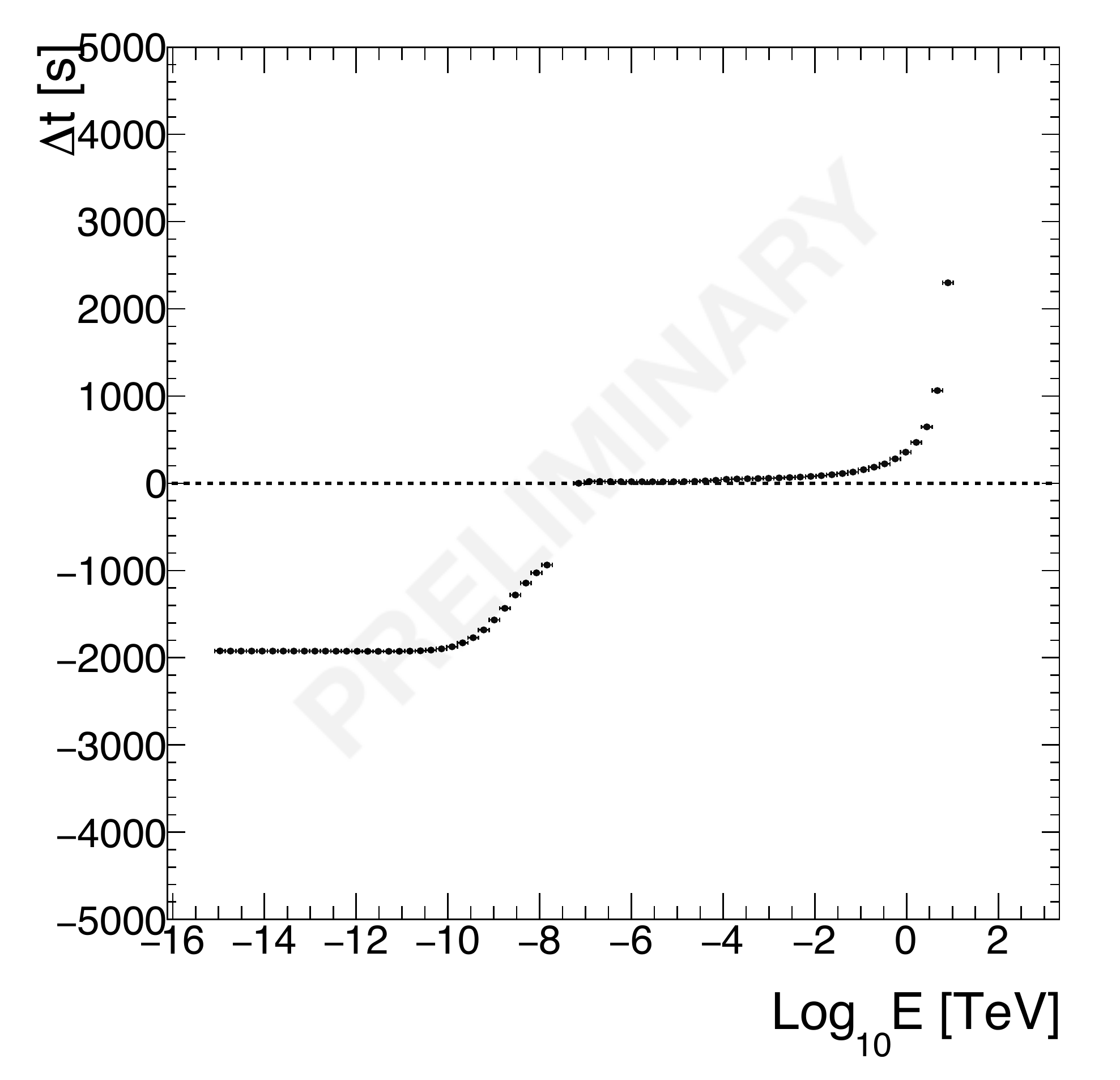}
\vspace{-0.35in}
\subcaption{$B_0=5$ mG (increasing).}
\label{fig:incr}
\end{minipage}
\vspace{-0.1in}
\caption{Intrinsic delays extended towards a lower energy domain, from IR to gamma-ray ranges.} 
\label{fig:delays}
\end{figure}

We extend the previously derived intrinsic delays from X-rays down to infrared energies. To better identify and separate both energy domains, a special care is given to the choice of the reference arrival time $t_{ref}$ at $E_{ref}$. We use the coordinates of SEDs crevice between the two bumps to specify this quantity (located at $\log_{10} E_{(\mathrm{TeV})} \sim -8$ in Figure~\ref{fig:SED}). Since the location of the crevice varies between all the evolving SEDs, we only consider the one with the highest flux value $\nu F_{\nu}$ as reference. The energy $E_{ref}$ is then taken at the crevice energy for that specific SED. $E_{ref}$ usually varies between 0.01 and 1 MeV. To highlight the relationship between the two energy domains, we vary one source parameter to send delays into another regime, as shown in Figure~\ref{fig:delays}. A systematic similarity of time delays in the two energy domains appears. Both sets of delays always follow the same trend whichever set of parameters is used to generate them. To fully exploit this potential we attempt a better estimation of the degree of similarity with euclidian distance.

\section{Euclidian distance on time delays}
The euclidian distance between two data sets $A[i]_{(1 \leqslant i \leqslant N)}$ and $B[i]_{(1 \leqslant i \leqslant N)}$ containing $N$ points is simply written as:
\begin{equation}
\label{eq:eucl_scalar}
d_{E} = \frac{\sqrt{\sum_i \left(A[i]-B[i]\right)^2}}{\sqrt{\sum_i \left(A[i]+B[i]\right)^2}},
\end{equation}
where the denominator is a normalisation term we introduce for a better readability. Two values stand out as noteworthy identities:
\begin{itemize}[noitemsep]
\item $d_{E}=0$ is the minimum indicating a perfect match between the two data sets, i.e. $A = B$;
\item $d_{E}=1$ either for $A[i]=0$ or $B[i]=0$ for all $i$. 
\label{eq:eucl}
\end{itemize}

We apply a treatment inspired from signal processing zero-padding and create two data sets: one that contains synchrotron delays (with IC delays set to zero), and the other containing IC delays (with synchrotron delays set to zero). Both sets of delays are extended between $\log_{10}E_{(\mathrm{TeV})} = -16$ to $\log_{10}E_{(\mathrm{TeV})} = 2$, filling the areas where there is no information with zeros. To maximise the sensitivity of the above formula, we rescale all the data points with the first non zero value $\Delta t_{\mathrm{first}}$: $\Delta t \rightarrow \Delta t - \Delta t_{\mathrm{first}}$.

\begin{figure}
\vspace{-0.15in}
\begin{minipage}[t]{0.5\textwidth}
\centering
\includegraphics[width=0.66\linewidth]{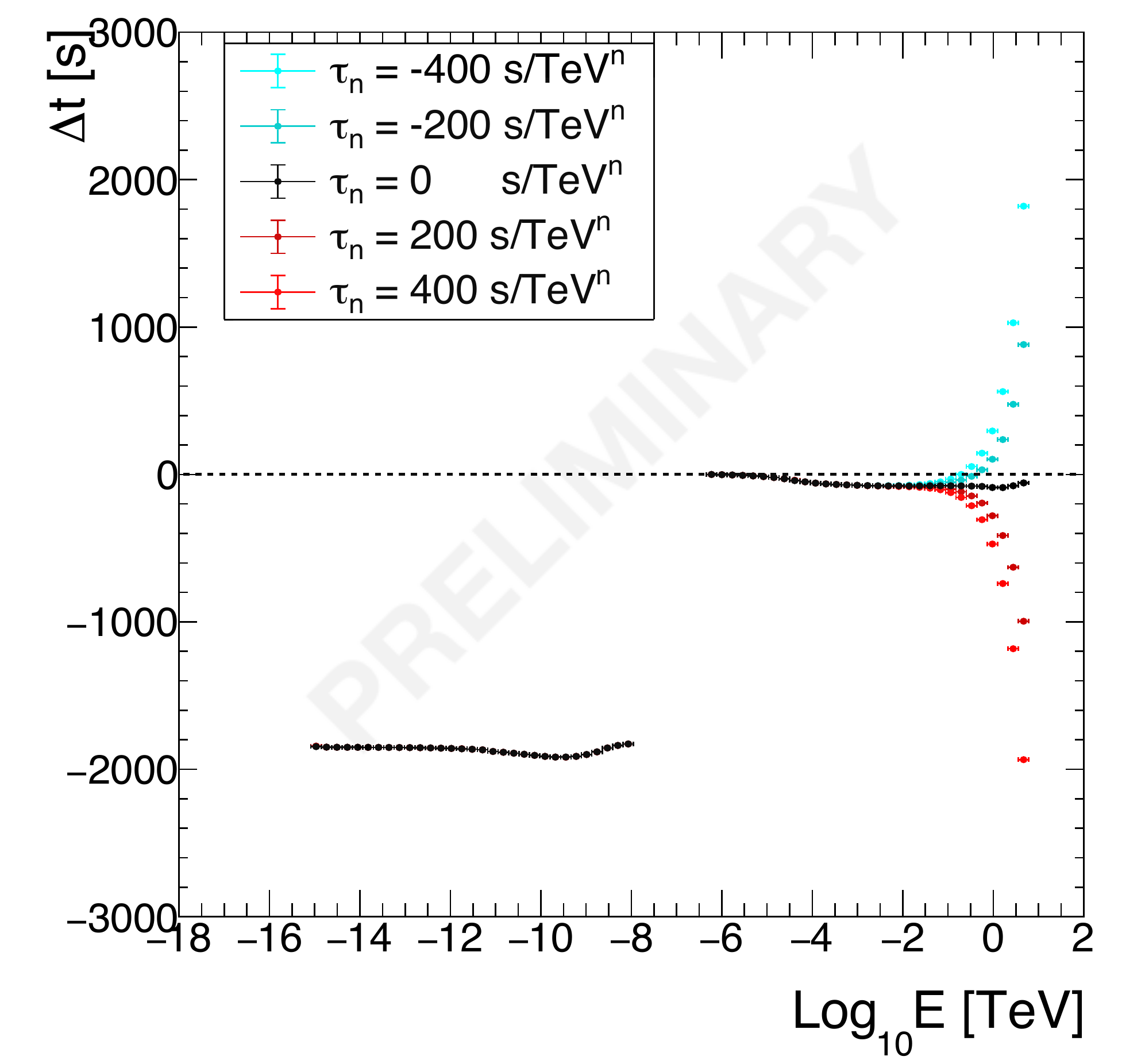}
\vspace{-0.15in}
\subcaption{Time delays.}
\label{fig:LIVevol_flat}
\end{minipage}\hfill
\begin{minipage}[t]{0.5\textwidth}
\centering
\includegraphics[width=0.66\linewidth]{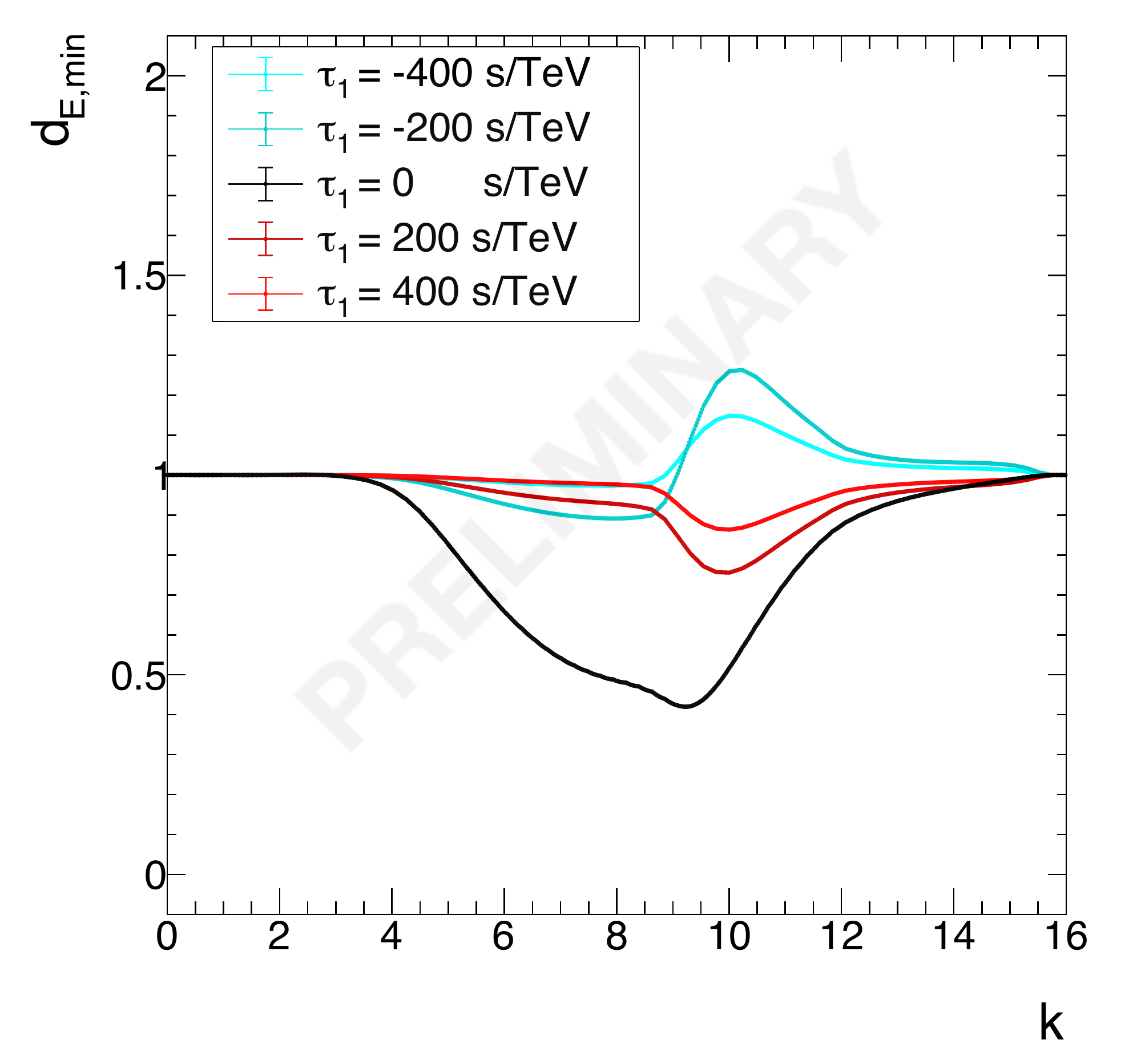}
\label{LIVEDevol_flat}
\vspace{-0.15in}
\subcaption{Euclidian distance.}
\label{fig:LIVEDevol_flat}
\end{minipage}
\vspace{-0.15in}
\caption{Time delays and euclidian distance function computed for the flat regime with various LIV contribution. The minimum euclidian distance $d_{E,min}$ is reached by the curve with no LIV effect (black).}
\vspace{-0.15in}
\label{fig:ED}
\end{figure}

To ensure a good overlap we shift one of the data sets over energy (i.e. horizontally on Figure~\ref{fig:LIVevol_flat}) towards the other one and find the displacement $k$ that minimises the euclidian distance $d_{E}$. We thus introduce a modified version of equation \ref{eq:eucl}, where $d_{E}$ now becomes a function of the displacement $k$:
\begin{equation}
\label{eq:eucl_k}
d_{E}[k] = \frac{\sqrt{\sum_i \left(A[i-k]-B[i]\right)^2}}{\sqrt{\sum_i \left(A[i-k]+B[i]\right)^2}},
\end{equation}
with $A$ being the displaced data set. For our study, we compute the distance in logarithmic scale by displacing the synchrotron data set ($\equiv A$) over energy by a quantity $\epsilon = 10^k$ towards the IC set ($\equiv B$) such that $E_{new} = E\times 10^k$. Although our data sets are discrete, we treat the distributions of time delays as functions and retrieve values with linear interpolations between consecutive points such that $k$ can take any value. When time delays present strong variations, the minimum of the euclidian distance function indicates the optimal displacement $k_{min}$ to apply to one set of delays to reach a best match with the other set. The accuracy on the reproduction is then indicated by the minimal distance $d_{E,min}$.

The extended time delays and euclidian distance between the synchrotron and IC data sets obtained with LIV effects of various values are shown in Figure~\ref{fig:ED}. This is a very specific case with little to no variation in the rather flat delays, for which the sensitivity of this method is the lowest. However one can easily see the LIV contribution tends to decorrelate synchrotron and IC sets of delays. The distance function minimum $d_{E,min}$ shifts away from 0 as $|\tau_n|$ increases. Generally speaking, SSC flares always correspond to $d_{E,min} < 0.6$ while introducing LIV effects results in equal or higher values of $d_{E,min}$. \textit{Values of $d_{E,min}>0.6$ are therefore characteristic of the presence of non-intrinsic time delays.}

\section{Hysteresis pattern on SEDs}
The similarity between synchrotron and IC domains can also be quantified directly from SEDs. Hardness-intensity diagrams (HID) showing the SED index against the SED flux give rise to hysteresis patterns. Such patterns have been found in observed data \cite{Hystobs1,Hystobs2,Hystobs3} and simulated with flare models \cite{Hystexp1}, and ultimately help to constrain emission scenarii. We produce such diagrams by computing the hardness over a small energy window in both synchrotron and IC domain, the same for all SEDs. To account for LIV effects, we rebuild SEDs using LIV-shifted light curves. Contrarily to time delays, LIV contribution to SEDs is barely noticeable but they can be highlighted through HIDs. The intensity plotted in the diagram is taken as the mean flux within the energy window chosen such that there is no overlap with SED maxima to prevent a change of sign in spectral indices. We thus define the synchrotron and IC windows between the position of the two peaks at the highest energy (i.e. far right peak) and spread the windows over one decade to higher energies. In other words, we focus here over one decade energy windows (X-ray and gamma-ray energy bands) on the high-energy side of each peak.

\begin{figure}
\vspace{-0.25in}
\begin{minipage}[t]{0.32\textwidth}
\centering
\includegraphics[width=1\linewidth]{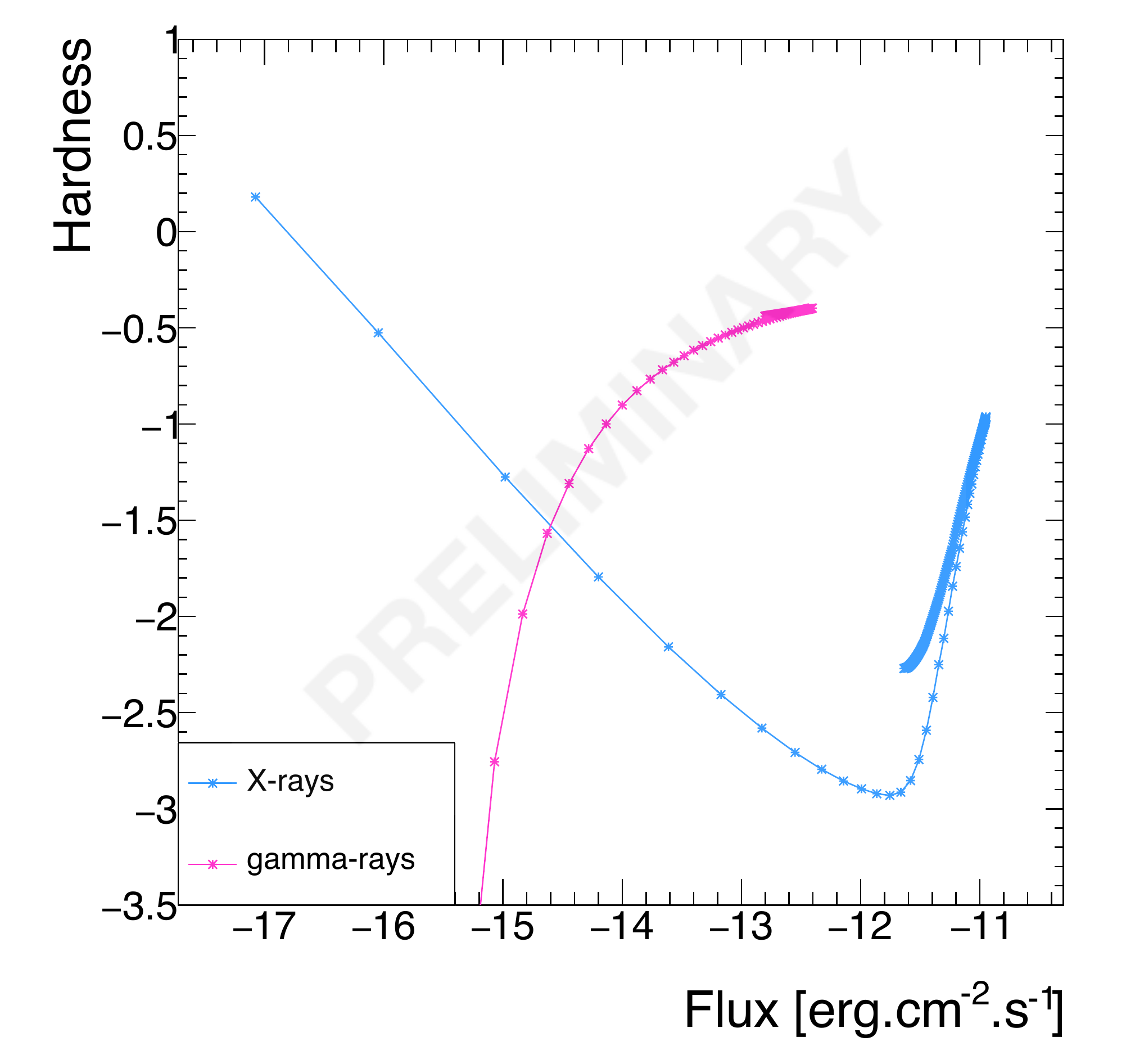}
\vspace{-0.25in}
\subcaption{$\tau_1 = 0$ s/TeV.}
\label{fig:Hyst_flat}
\end{minipage}\hfill
\begin{minipage}[t]{0.32\textwidth}
\centering
\includegraphics[width=1\linewidth]{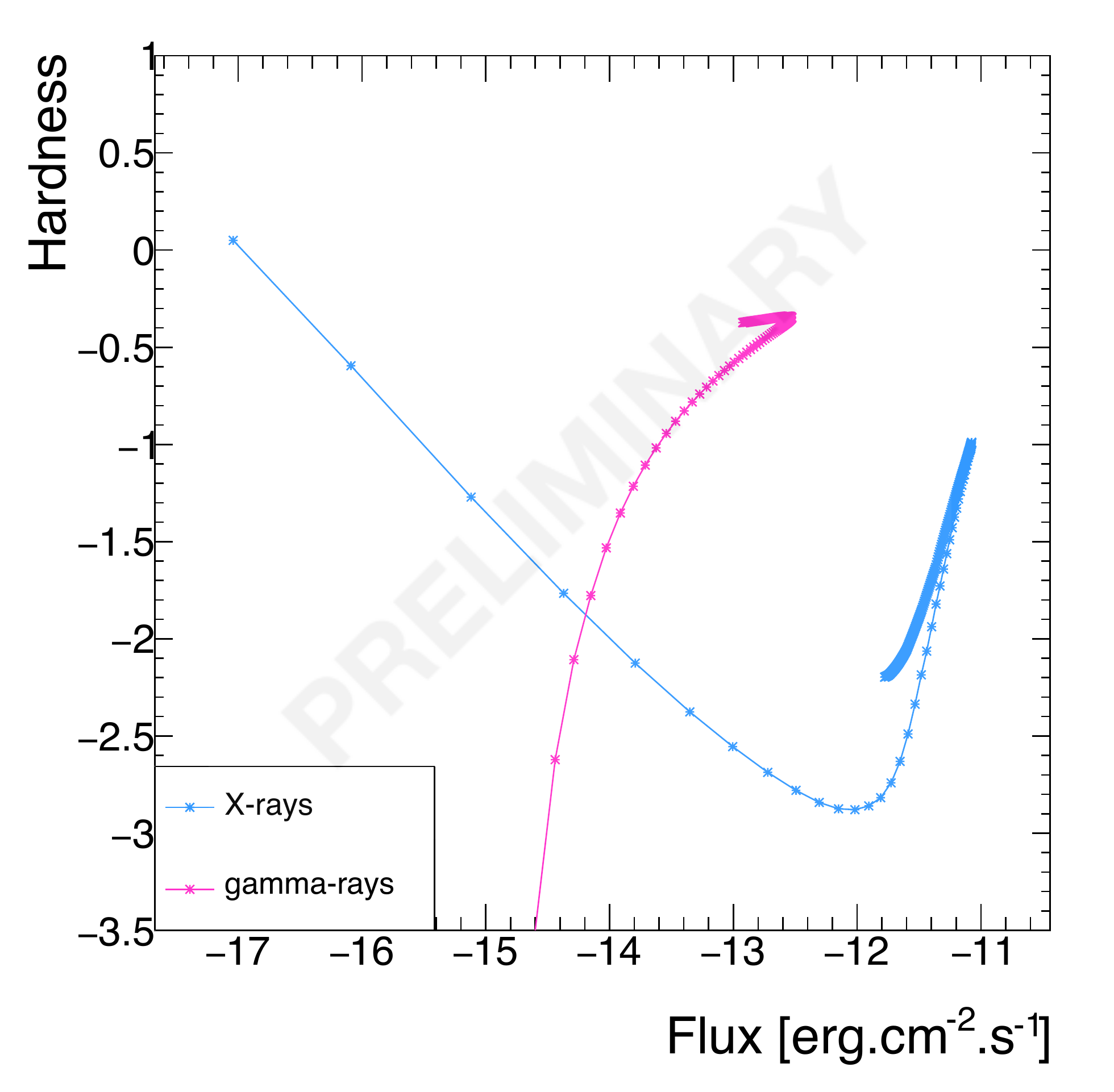}
\vspace{-0.25in}
\subcaption{$\tau_1 = 400$ s/TeV.}
\label{fig:Hyst_flatPos}
\end{minipage}\hfill
\begin{minipage}[t]{0.32\textwidth}
\centering
\includegraphics[width=1\linewidth]{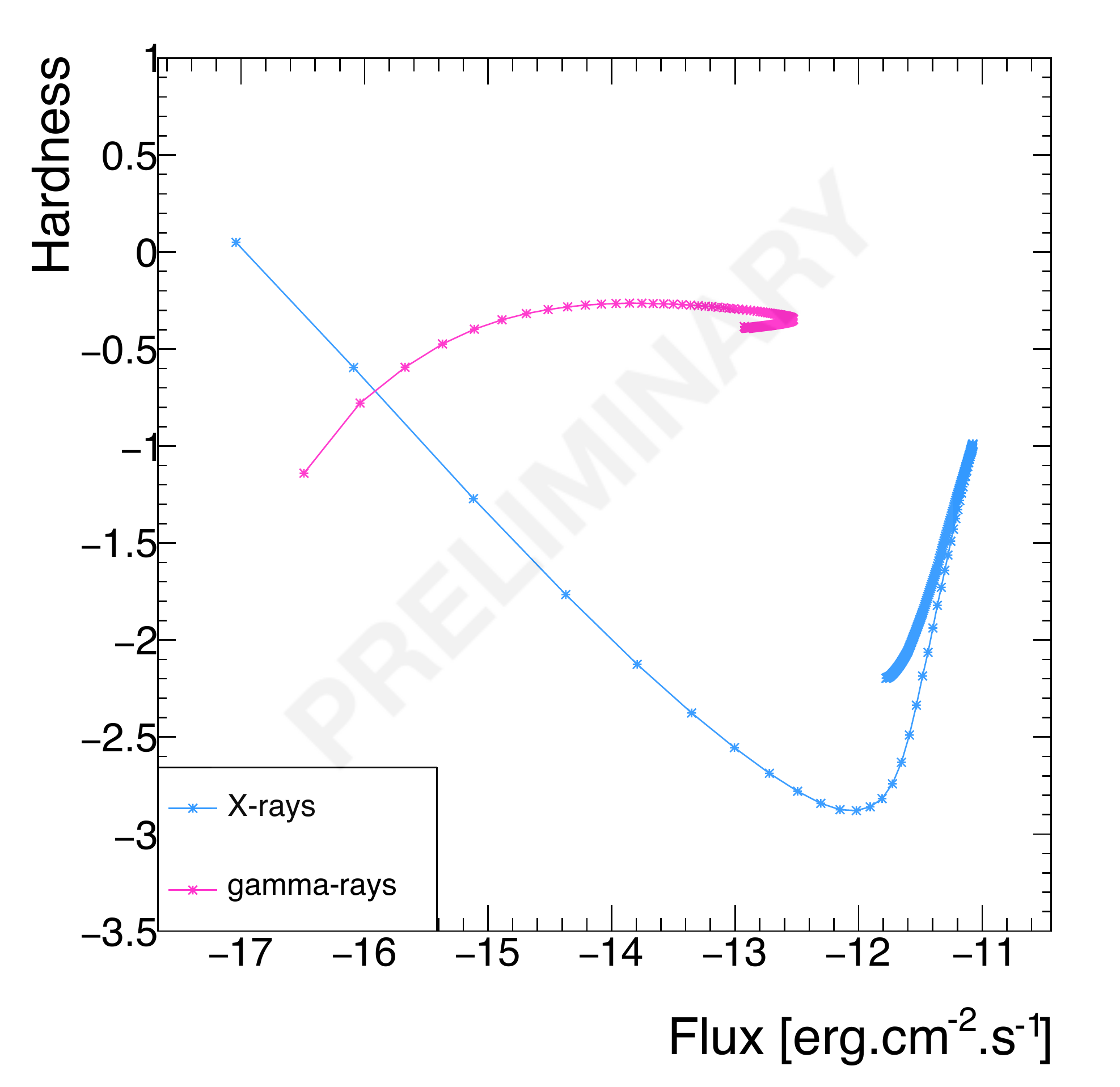}
\vspace{-0.25in}
\subcaption{$\tau_1 = -400$ s/TeV.}
\label{fig:nHyst_flatNeg}
\end{minipage}
\caption{Hardness-intentity diagrams evolution for the flat regime with various LIV contribution showing LIV effects can cause a change in loop orientation in the gamma-ray domain.}
\vspace{-0.15in}
\label{fig:Hyst}
\end{figure}

Figure~\ref{fig:Hyst} shows the resulting hysteresis pattern for the flat regime with various LIV effects. When $\tau_n=0$, similarly to time delays, hysteresis patterns are the same for both domains. X-rays and gamma-rays hysteresis always follow the same loop orientation which changes according to the delay regime. Increasing (resp. decreasing) regime leads to anti-clockwise (resp. clockwise) loop orientation, and flat regime leads to patterns with only weak hysteresis. For  $\tau_n \neq 0$, the hysteresis for X-ray energies (blue) is unaffected by LIV and provides a reference to be compared with gamma-ray hysteresis. One then observes the following correspondence in gamma-rays:
\begin{itemize}[noitemsep]
\item $\tau_n>0 \Longleftrightarrow$ anti-clockwise orientation $\Longleftrightarrow$ increasing delays;
\item $\tau_n<0 \Longleftrightarrow$ clockwise orientation $\Longleftrightarrow$ decreasing delays.
\end{itemize}

Large enough $\tau_n$ can cause hysteresis to change orientation. The information on the X-ray domain hysteresis still indicates the regime of intrinsic delays at work in the gamma-ray domain. Alternatively, \textit{witnessing a difference of looping orientation between the X-ray and gamma-ray hysteresis patterns strongly hints at another contribution being at play.}  This tool can be used to resolve the existence of intrinsic time delays even though none is detected due to a lack of precision or because another phenomenon (such as LIV effects) contributes to cancelling them out.

\section{Discussion and conclusion}
Despite the EBL and Klein-Nishina effects that reduce the flux at the highest energies, a strong correlation arises between observables in the synchrotron and IC domains in SSC scenarii. Euclidian distance appears as an interesting tool allowing us to deduce intrinsic delays in the gamma-ray energies from the ones in the X-ray domain. Furthermore, we have identified a threshold where time delays can no longer be explained by intrinsic effects in a pure SSC scenario which provides a way to identify the presence of non-intrinsic delays, such as the ones due to LIV. Any deviation from a standard pure SSC flare, such as additional external inverse Compton or adiabatic processes, can reduce the symmetry between synchrotron and gamma-ray properties. Alternatively the presence of such a symmetry could help validate the pure one zone SSC scenario.

Anyways, introducing LIV effects in our simulations leads to a significant break of the symmetry. This asymmetry can however be used to our advantage. Indeed in the eventuality where the pure one zone SSC scenario is validated with other observables, a break of symmetry would hint that another effect is at play and contributes to modify the time delays. In the case where instruments sensitivity allows it, this can also be further confirmed with the study of hysteresis where a similar break of symmetry should arise. We may thus be able to discriminate between intrinsic and LIV contributions to observed time delays with such tools and methods.

%
%
%

\end{document}